\documentclass[aps,prb,twocolumn,floatfix,superscriptaddress,showpacs]{revtex4}
\usepackage{graphicx}

\usepackage{graphicx}
\usepackage{amssymb}
\usepackage{ulem}

\begin{document}

\title{Surface electron mobility over a helium film}
\author{D\'{e}bora Coimbra}
\affiliation{Departamento de F\'{\i}sica, Universidade Federal de S\~ao Carlos, \\
13565-905 S\~ao Carlos SP, Brazil}
\author{Sviatoslav S. Sokolov}
\affiliation{B. I. Verkin Institute for Low Temperature Physics and Engineering,\\
National Academy of Sciences of Ukraine, 61103 Kharkov, Ukraine}
\author{J.-P. Rino}
\affiliation{Departamento de F\'{\i}sica, Universidade Federal de S\~ao Carlos, \\
13565-905 S\~ao Carlos SP, Brazil}
\author{Nelson Studart}
\affiliation{Departamento de F\'{\i}sica, Universidade Federal de S\~ao Carlos, \\
13565-905 S\~ao Carlos SP, Brazil}
\date{\today }

\begin{abstract}
The mobility of surface electrons localized over helium films
underlying solid substrates has been evaluated by solving the
Boltzmann equation in the time relaxation approximation and the
force balance equation in which an effective mobility is obtained
in terms of the dynamical structure factor of the nondegenerate
electron liquid. The essential processes of electron scattering by
gas atoms, ripplons, and film-solid interface roughness are taken
into account. The electron mobility dependence on the film
thickness and temperature is determined and compared with
experimental data available. We find that the interface-roughness
scattering is the dominant process for explaining the experimental
results. We estimate the extended defect sizes of the underlying
substrate within the Gaussian correlated model for interface
roughness.
\end{abstract}

\maketitle

\section{Introduction}

Surface electrons (SE) levitating over liquid helium has been used
as a paradigmatic quasi-two dimensional electron system (Q2DES).
It constitutes the counterpart of the electron system on
semiconductor heterostructures in the sense that the electrons on
helium obey the classical statistics because the low densities
achievable in experiments ($\lesssim 10^{9}$ cm$^{-2}$). One point
difference between them comes from the scattering mechanisms. In
the former system, electrons are scattered by surface excitations,
the ripplons, and by vapor atoms of the helium surface. In the
latter one, scattering by impurities, phonons, and interface
roughness are the processes that limit the electron mobility.
Besides the remarkable phenomena discovered in this
low-dimensional electron system, for instance, Wigner
crystallization\cite{grimes79} and the existence of
edgemagnetoplasmons\cite{mast84,glattli84} SE have been used as a
probe to investigate the elementary surface excitations of
cryogenic liquids and solids. Very recently there is an intensive
search for experimental realization of SE as qubits leading to a
quantum computer.\cite{dahm03,kono02,papa05} For a overview of the
field see Ref. \onlinecite{review97}.

For electrons trapped on the bulk helium surface it is well known
that the SE mobility is dominated by the electron-helium gas atom
scattering for $T>1$ K whereas the SE scattering by quantized
capillary waves (ripplons) is responsible for mobility at lower
temperatures where the helium vapor density becomes negligible.
The experimental data for SE mobility over bulk helium are in
reasonable agreement with theoretical calculations.\cite{iye80}

The situation changes when SE are floating over a helium film
which in turn is deposited over a solid substrate. In such a
condition it is possible to
increase significantly the accessible range of electron densities\cite%
{mistgunz97} and the electron correlations may become important.
Furthermore, the electron energy spectrum, the ripplon dispersion
relation and the electron-ripplon interaction are modified
significantly due to film effects and the van der Waals forces
from the substrate now play a decisive role in the transport
properties. Furthermore, besides the usual scattering mechanisms
pointed out above, we must consider the SE scattering by interface
defects at the helium film - substrate boundary, which, as it have
been shown in experiments is responsible by the unusual behavior
of the SE transport
properties on a helium film.\cite%
{platzman86,jiangdahm87,jiangstandahm88,hucarmidahm92,dahm97}

The present work intends to provide a detailed theoretical study
of the mobility of SE localized over a helium film deposited on
substrates, as solid neon, glass and poly(methyl-methacrylate)
[pmma], which are the materials that have been used in the
experiments. We employ the Boltzmann transport equation approach
(BEA) and the force balance equation method (FBEM) in which
Coulomb effects on the mobility are taken into account through the
dynamical structural factor of the Q2DES. We calculate the SE
mobility as a function of temperature and the film thickness in
the range $100$\ $<d<$ $1000$ \AA . From the comparison with
experimental results we check the role of all scattering processes
involved and verify the reliability of the interface roughness
model used in our treatment. We find that the SE mobility is
strongly determined by the interface roughness substrate and the
results are in quite good agreement with the experiments.

\section{Basic Relations}

\subsection{Surface electron states}

Electron states on a liquid helium film are confined in the
direction normal to the surface ($z$) due to the potential well
created by a infinitely high barrier which prevents the electrons
penetrate inside the liquid phase and by and attractive potential
due to the electron interaction with the polarizable substrates
and an applied holding electric field $E_{\perp }$ along the $z$
direction. The electron potential energy over flat helium film
located at $-d<z<0$ is well known and given by\cite{shikmon74}%
\begin{equation}
U(z)=-\frac{\Lambda _{0}}{z}-\frac{\Lambda _{1}}{z+d}+eE_{\perp }
\label{1}
\end{equation}%
where $\Lambda _{0}=e^{2}(\varepsilon _{He}-1)/4(\varepsilon _{He}+1)$ and $%
\Lambda _{1}=e^{2}\varepsilon _{He}(\varepsilon _{s}-\varepsilon
_{He})/\left( \varepsilon _{He}+1\right) ^{2}(\varepsilon
_{s}+\varepsilon _{He})$ with $\varepsilon _{He}$ \ and \
$\varepsilon _{s}$ being the dielectric constants of helium and
solid substrate, respectively, and $e$ the electron charge.

For electrons moving freely along the helium surface with the wave
function
and the energy spectrum given respectively by $\psi _{l}(\mathbf{r}%
,z)=A^{-1/2}\exp (i\mathbf{k\cdot r})\chi _{l}(z)$ and
$E_{l}(k)=\hslash ^{2}k^{2}/2m+\Delta _{l}$, where $\mathbf{k}$
and $\mathbf{r}$ are the wave
vector and position vector in the plane $(x,y)$ of the liquid interface and $%
A$ is surface area occupied by electrons. Unfortunately an
analytical solution of Schr\"{o}dinger equation for $\chi _{l}(z)$
and subbands
energies $\Delta _{l}$ cannot be found for $U(z)$ given by Eq. (%
\ref{1}). In this work, we use the variational method by choosing
the trial
wave function $\chi _{1}(z)=2\gamma ^{3/2}ze^{-\gamma z}$. The parameter $%
\gamma $\ depends strongly on $d$ and is determined from the
subband energy minimization as was done in Refs.
\onlinecite{sokrinostud97} and \onlinecite{sokstud03modes}. One
has found that the energy gap between the ground and first excited
subbands increases considerably by decreasing $d$. For example
$\Delta _{2}-\Delta _{1}\simeq 12.7$ K for $d=10^{-6}$ cm for the
neon substrate and is much larger for a substrate with higher
$\varepsilon _{s}$. \cite{sokstud03modes} Then one can disregard
the possibility of electron transition to excited subbands which
is proportional to $\exp [-(\Delta _{2}-\Delta _{1})/T]$ and take
only the ground subband into account in the calculation of the
scattering matrix elements.

\subsection{Interaction Hamiltonians}

The electron interaction with helium vapor atoms by a contact type
Hamiltonian is given by
\begin{equation}
\widehat{H}_{eg}=\frac{2\pi \hslash ^{2}a_{0}}{m}\sum_{e,a}\delta (\mathbf{R}%
_{e}-\mathbf{R}_{a})  \label{3}
\end{equation}%
where $\mathbf{R}_{e}$ and $\mathbf{R}_{a}$ are three-dimensional
positions of electrons and atoms, respectively and $a_{0}\simeq
0.61$ \AA\ is the scattering length.\cite{saitoh77} The
electron-ripplon interaction was derived in Ref.
\onlinecite{shikmon74} and the electron interaction with surface
roughness was obtained in the case of SE over solid hydrogen.\cite%
{sokrinostud95}

Very recently two of us have constructed an Hamiltonian where both
electron-ripplon and electron-solid interface interactions are
treated on the same footing.\cite{sokstud03} The resulting total
interaction potential for any substrate was obtained from the
solution of the Poisson's equation through the perturbation of the
boundary conditions from the flat positions at $z=0$ and $z=-d.$
Due to very small polarizability of liquid helium, the result of
our approach for the electron-ripplon interaction coincides with
that obtained in Ref. \onlinecite{shikmon74}. Bearing this in
mind, we write the Hamiltonian for the electron-ripplon and
electron-interface couplings in a unique form as
\begin{equation}
\widehat{H}_{er(ei)}=\frac{1}{\sqrt{A}}\sum_{\mathbf{q}}\xi _{1(2)\mathbf{q}%
}V_{r(i)\mathbf{q}}(z)e^{i\mathbf{q\textbf{.}r}},  \label{4}
\end{equation}%
with\cite{shikmon74}
\[
V_{r\mathbf{q}}(z)=\frac{\Lambda _{0}q}{z}\left[ \frac{1}{qz}-K_{1}(qz)%
\right] +\frac{\Lambda _{1}}{\left( z+d\right) ^{2}}+eE_{\bot }
\]%
and\cite{sokstud03}

\begin{eqnarray*}
V_{iq}(z) &=&-\Lambda _{1}\left\{ \left( \frac{2\varepsilon _{He}}{%
\varepsilon _{s}+\varepsilon _{He}}\right) \frac{qK_{1}\left[
q\left(
z+d\right) \right] }{(z+d)}\right.  \\
&&\left. +\left( \frac{\varepsilon _{s}-\varepsilon
_{He}}{\varepsilon _{s}+\varepsilon _{He}}\right)
\frac{q^{2}K_{2}[q(z+d)]}{2}\right\} ,
\end{eqnarray*}%
where $K_{j}(x)$ is the modified Bessel function. Here we used the
Fourier
transform $\xi _{j}\left( \mathbf{r}\right) =A^{-1/2}\sum_{\mathbf{q}}\xi _{j%
\mathbf{q}}e^{i\mathbf{q.r}}$ for $\xi _{1}\left( \mathbf{r}\right) $ and $%
\xi _{2}\left( \mathbf{r}\right) $ and, quantizing the
oscillations of free
helium surface, one obtains%
\begin{equation}
\xi _{1\mathbf{q}}=\sqrt{\frac{\hbar q\tanh (qd)}{2\rho \omega
_{q}}}\left( a_{\mathbf{q}}+a_{-\mathbf{q}}^{\dagger }\right)
\label{5}
\end{equation}%
with\ $a_{\mathbf{q}}$\ and\ $a_{\mathbf{q}}^{\dagger }$\ being
the
annihilation and creation operators for ripplons with\ wave number\ $\mathbf{%
q}$ and satisfying the dispersion law
\begin{equation}
\omega _{q}^{2}=\left( \frac{\alpha }{\rho }q^{3}+g^{\prime
}q\right) \tanh qd,  \label{6}
\end{equation}%
where\ $\alpha $ is the surface tension coefficient and $\rho $ is
the density of helium. In\ Eq. (\ref{6}),\ $g^{\prime }=g+3\beta
/\rho d^{4},$\ with $g$ being the gravity acceleration and $\beta
$\ the van der\ Waals constant.

Even though one cannot find from first principles the interface
roughness displacement $\xi _{2}\left( \mathbf{r}\right) $, we
will use a reasonable model to calculate the interface scattering
contribution to SE transport properties.

\section{Transport properties}

\subsection{Boltzmann kinetic approach}

In order to calculate SE mobility along the helium surface in the
presence of a driving electric field $\mathbf{E}_{\parallel }$ we
apply the well-known kinetic formalism where the electron
scattering plays a crucial role in transport properties and does
contribute to collision integrals in the Boltzmann equation as
follows \
\begin{equation}
\frac{e\mathbf{E}_{\parallel }}{\hbar }\frac{\partial f}{\partial \mathbf{k}}%
=\widehat{S}_{eg}\left\{ f\right\} +\widehat{S}_{er}\left\{ f\right\} +%
\widehat{S}_{ei}\left\{ f\right\} ,  \label{7}
\end{equation}%
where\ $f(\mathbf{k})$ is the electron distribution function and \ $\widehat{%
S}_{eg}\left\{ f\right\} $, \ $\widehat{S}_{er}\left\{ f\right\} $, and $%
\widehat{S}_{ei}\left\{ f\right\} $ are collision integrals\ of
electron with helium atoms,\ ripplons and\ roughness interface
between the film and the substrate,\ respectively. By considering
quasielastic scattering processes the solution of Eq. (7) is
$f(\mathbf{k})\simeq f^{(0)}(k)+f^{(1)}(k)\cos \varphi $ where
$f^{(0)}(k)$ is equilibrium
distribution given by the Boltzmann function for the nondegenerate Q2DES, $%
\varphi $ is the angle between $\mathbf{k}$ and
$\mathbf{E}_{\parallel }$, and
\[
f^{(1)}(k)=-\frac{e\mathbf{E}_{\parallel }}{\hslash \nu
(k)}\frac{\partial f^{(0)}(k)}{\partial \mathbf{k}},
\]%
where the collision frequency $\nu (k)=\nu _{eg}(k)+\nu
_{er}(k)+\nu _{ei}(k) $ and
\begin{equation}
\nu _{j}(k)=\sum\limits_{\mathbf{k}^{\prime }}W_{j}(\mathbf{k}^{\prime },%
\mathbf{k})\left( \frac{1-f_{0}(k^{\prime })}{1-f_{0}(k)}\right)\left[ 1-\frac{\mathbf{k}%
^{\prime }\textbf{.}\mathbf{k}}{k^{2}}\right] .  \label{8}
\end{equation}%
The probability amplitude per unit-time
$W_{j}(\mathbf{k},\mathbf{k}^{\prime })$ for electron transition
from states $\mathbf{k}$ to\ $\mathbf{k}^{\prime }$ is given by
Fermi's golden rule and the subscript $j$ corresponds to each
scattering mechanism whose interaction potential is given by Eqs.
(\ref{3})
and (\ref{4}). The SE mobility is then given by%
\begin{equation}
\mu =\frac{e}{m}\int\limits_{0}^{\infty }\frac{xe^{-x}dx}{\nu
\left( k_{T}x^{1/2}\right) }  \label{9}
\end{equation}%
where $x=k^{2}/k_{T}^{2}$, with the thermal wave vector $k_{T}=\sqrt{2mT}%
/\hslash .$

The frequency of electron-atom collisions has been calculated in Ref. \onlinecite%
{saitoh77} and the result is
\begin{equation}
\nu _{eg}=\frac{3\pi ^{2}a_{0}^{2}\hbar n_{g}\gamma }{8m}.
\label{10}
\end{equation}%
Note that $\nu _{eg}$ does not depend on $k$ being a function of
volume concentration of helium atoms $n_{g}=(MT/2\pi \hbar
^{2})^{3/2}\exp (-Q/T),$ where $Q$ is the vaporization energy and
$M$\ is the $^{4}$He mass. As $n_{g}$ is an increasingly
exponential function of $T,$ $\nu _{eg}$ becomes negligible in
comparison with $\nu _{er}$ and $\nu _{ed}$ for $T<1$ K.

The calculation of $\nu _{er}$ and $\nu _{ei}$ is more
complicated. Using Eqs. (\ref{4}) and (\ref{8}) one can obtain in
a straightforward way that both $\nu _{er}(k)$ and $\nu _{ei}(k)$
can be written as
\begin{equation}
\nu _{er(i)}\left( k\right) =\frac{m}{\pi \hbar ^{3}k^{2}}%
\int\limits_{0}^{2k}\frac{q^{2}<\left\vert \xi
_{1(2)\mathbf{q}}\right\vert ^{2}>\left\vert \langle \chi
_{1}|V_{r(i)q}(z)|\chi _{1}\rangle \right\vert
^{2}}{\sqrt{4k^{2}-q^{2}}}dq,  \label{12}
\end{equation}%
where $\langle ...\rangle $ means an average over ensemble.
Considering the electron-ripplon scattering one obtains, from Eq.
(\ref{5}),

\begin{equation}
\left\langle \left\vert \xi _{1(2)\mathbf{q}}\right\vert ^{2}\right\rangle =%
\frac{\hslash q\tanh (qd)}{2\rho \omega _{q}}(2N_{q}+1)
\label{13}
\end{equation}%
where the ripplon number $N_{q}$ is given by the Bose-Einstein
function, which leads to \ $2N_{q}+1=\coth \left( \hbar \omega
_{q}/2T\right) \simeq 2T/\hbar \omega _{q}\gg 1$ for long
wavelength ripplons with $\hbar \omega _{q}\ll T$. Equation
(\ref{12}) results into
\begin{equation}
\nu _{er}(k_{T}x^{1/2})=\nu _{er}^{(0)}J(x)
\label{14}
\end{equation}%
where $\nu _{er}^{(0)}=8\gamma _{0}^{2}T^{2}/\pi \hbar \alpha $ and%
\begin{eqnarray}
J(x) &=&x^{2}\int\limits_{0}^{\pi /2}\Phi _{er}^{2}\left( \sqrt{\frac{xT}{%
\Delta }}\sin \theta \right) \frac{\sin ^{6}\theta d\theta }{(\sin
^{2}\theta +x_{c}/x)}+  \nonumber \\
&&\frac{2}{3}x\sqrt{\frac{\Delta _{\perp }^{3}}{\Delta _{0}T^{2}}}%
\int\limits_{0}^{\pi /2}\Phi _{er}\left( \sqrt{\frac{xT}{\Delta
}}\sin \theta \right) \frac{\sin ^{4}\theta d\theta }{(\sin
^{2}\theta +x_{c}/x)}
\nonumber \\
&&+\frac{\pi }{18}\frac{\Delta _{\perp }^{3}}{\Delta _{0}T^{2}}\left[ 1-%
\frac{x_{c}/x}{\sqrt{1+\left( x_{c}/x\right) ^{2}}}\right] .
\end{eqnarray}%
Here
\begin{eqnarray}
\Phi _{er}\left( y\right)  &=&\phi (y)+\frac{\Lambda _{1}}{\Lambda _{0}y^{2}}%
\left[ 1+2\gamma d+\right.   \nonumber \\
&&\left. 4\gamma d\left( 1+\gamma d\right) e^{2\gamma d}{Ei}\left(
-2\gamma d\right) \right] ,  \label{16}
\end{eqnarray}%
with \
\begin{eqnarray*}
\phi (y) &=&-\frac{1}{1-y^{2}} \\
&&-\left[ \frac{1}{\sqrt{\left( 1-y^{2}\right) ^{3}}}\ln \left( \frac{y}{1+%
\sqrt{1-y^{2}}}\right) \right] \Theta (1-y) \\
&&-\left[ \frac{1}{\sqrt{\left( y^{2}-1\right) ^{3}}}\arcsin \left( \frac{%
\sqrt{y^{2}-1}}{y}\right) \right] \Theta (y-1),
\end{eqnarray*}%
\ In the above expressions $\Delta =\hbar ^{2}\gamma ^{2}/2m,$\
$\Delta _{\perp }=\hbar ^{2}\gamma _{\bot }^{2}/2m,$\ $\Delta
_{0}=m\Lambda
_{0}^{2}/2\hslash ^{2},\ \gamma _{\bot }=(3meE_{\bot }/2\hslash ^{2})^{1/3}$%
, $x_{c}=\hbar ^{2}\rho g^{\prime }/8m\alpha T$,  $Ei(y)$ is the
exponential-integral function, and $\Theta (y)$ is the step
function.

For the scattering by interface roughness, one obtains
\begin{eqnarray}
\nu _{ei}\left( k\right)  &=&\frac{16m\Lambda _{1}^{2}k^{4}}{\pi \hbar ^{3}}%
\int\limits_{0}^{\pi /2}\sin ^{6}\theta \;\langle \left\vert \xi
_{2q}(q=2k\sin \theta )\right\vert ^{2}\rangle   \nonumber \\
&&\times \Phi _{ei}^{2}\left( \frac{k}{\gamma }\sin \theta \right)
d\theta , \label{17}
\end{eqnarray}%
where\
\begin{equation}
\Phi _{ei}\left( y\right) =\frac{1}{y}\int\limits_{0}^{\infty }\frac{%
x^{2}K_{1}\left[ y\left( x+2\gamma d\right) \right] }{x+2\gamma
d}\exp (-x)dx.  \label{18}
\end{equation}%
In Eq. (\ref{18}) and thereafter $K_{j}(x)$ is the modified
$j$-order Bessel function. Now we must to choose an interface
roughness model to derive the amplitudes\ $\xi _{2q}$ appearing in
Eq. (\ref{12}). Following Prange and Nee \cite{prangenee78}, we
use the Gaussian correlated model, in which the
interface roughness is described by two characteristic sizes, the height\ $%
\xi _{0}$\ and the lateral length $l.$ These parameters define the
autocorrelation function for interface roughness which can be
written as \
\begin{equation}
\left\langle \xi \left( \mathbf{r}\right) \xi (\mathbf{r}^{\prime
})\right\rangle =\xi _{0}^{2}\exp \left( -\frac{|\mathbf{r}-\mathbf{r}%
^{\prime }|^{2}}{l^{2}}\right)   \label{19}
\end{equation}%
which leads to \
\begin{equation}
\left\langle \left\vert \xi _{2q}\right\vert ^{2}\right\rangle
=\pi \xi _{0}^{2}l^{2}\exp \left( -\frac{q^{2}l^{2}}{4}\right) .
\label{20}
\end{equation}%
Assuming this model, the collision frequency, given by Eq.
(\ref{17}), can be calculated and the result is\
\begin{equation}
\nu _{ei}\left( k_{T}x^{1/2}\right) =\nu _{ei}^{(0)}xF\left(
k_{T}x^{1/2}l\right) ,  \label{21}
\end{equation}%
where $\nu _{ei}^{(0)}=32m^{2}\Lambda _{1}^{2}\xi _{0}^{2}T/\hbar
^{5}$\ and\
\begin{equation}
F(b)=b^{2}\int\limits_{0}^{\pi /2}\sin ^{6}\theta \Phi _{ei}^{2}\left( \frac{%
b}{\gamma l}\sin \theta \right) \exp \left( -b^{2}\sin ^{2}\theta
\right) d\theta .  \label{22}
\end{equation}

The Gaussian correlated model was employed due to the successful
description of the interface roughness for similar systems, for
its simplicity and, since there are only two fitting parameters
$l$\ and $\xi _{0}.$

\subsection{Many-electron effects}

Monarkha and
co-workers\cite{vilkmon89,monarkha93,peters94,monteskewyder02}
have studied electron correlation effects in the transport as well
quantum magnetotransport of SE on helium by using the force
balance transport equation method (FBEM). In this approach the
frictional force experienced by the center of mass of the Q2DES
due to electron-scatterer interactions is evaluated through the
calculation of the momentum rate absorbed by the scatterers. For
the system of $N_{e}$ electrons the kinetic frictional force
$\mathbf{F}_{fr}$ is given by%
\begin{equation}
\mathbf{F}_{fr}=N_{e}e\mathbf{E}_{\parallel
}=-\frac{d\mathbf{P}}{dt}, \label{26}
\end{equation}%
where
\begin{eqnarray}
\frac{d\mathbf{P}}{dt} &=&-\frac{2\pi }{\hslash A}\left\langle
\sum_{\nu ^{\prime },j^{\prime }}\left( \mathbf{p}_{\nu ^{\prime
}}-\mathbf{p}_{\nu }\right) |\langle \nu ^{\prime },j^{\prime
}|\widehat{H}_{int}|\nu ,j\rangle
|^{2}\right.   \nonumber \\
&&\left. \delta \lbrack (E_{\nu ^{\prime }}+E_{j^{\prime
}})-(E_{\nu }+E_{j})]\right\rangle .  \label{27}
\end{eqnarray}%
Here $\mathbf{p}_{\nu }$ and $E_{\nu }$ are momentum and energy of
scatterer system in the state $|\nu \rangle $, $E_{j}$ the energy
of the electron system in the state $|j\rangle $,
$\widehat{H}_{int}$ is the Hamiltonian of the interaction between
the electron and the particular scatterer. The angle brackets
means a thermodynamic average in the laboratory coordinate system.
In Ref. \onlinecite{vilkmon89} is was shown that $d\mathbf{P}/dt$
can be written in terms of the dynamic form factor of the Q2DES

\[
S_{\text{lab}}(\mathbf{q},\omega )=\frac{2\pi \hslash
}{N_{e}}\left\langle \sum_{j^{\prime }}|\langle j^{\prime
}|n_{-\mathbf{q}}|j\rangle |^{2}\delta \left( E_{j^{\prime
}}-E_{j}-\hslash \omega \right) \right\rangle
\]%
in the laboratory frame. $S_{\text{lab}}(\mathbf{q},\omega )$ is
connected with the form factor $S(\mathbf{q},\omega )$ in the
center-of-mass of the electron system, moving at a velocity
$\mathbf{u}$ relative to the
laboratory frame, as $S_{\text{lab}}(\mathbf{q},\omega )=S(q,\omega -\mathbf{%
q\cdot u})$. Here $n_{\mathbf{q}}=\sum_{i}\exp (-i\mathbf{q\cdot
r}_{i})$. It is important to point out that electron correlations
are considered in the system through the many-particle dynamic
structure factor which is related to the density-density response
function $\chi (q,\omega )$ through the fluctuation-dissipation
theorem $S(q,\omega )=-(\hslash /2\pi n_{s})\coth (\hslash \omega
/2T)${Im}$\left[ \chi (q,\omega )\right] $, where $n_{s}$ is the
electron density.\cite{pinesnoz89}

In the limit of low velocities $\mathbf{u}$ satisfying the condition $%
\hslash \mathbf{q\cdot u}\ll T$, we may define an effective
collision frequency as a proportionality factor between the
momentum loss per unit time of the electron system and the average
electon velocity as
\begin{equation}
\frac{d\mathbf{P}}{dt}=-mN_{e}\widetilde{\nu }\mathbf{u.}
\label{28}
\end{equation}%
The effective collision frequency is the sum of frequencies for a
particular
interaction Hamitonian $\widetilde{\nu }=\widetilde{\nu }^{(eg)}+\widetilde{%
\nu }^{(er)}+\widetilde{\nu }^{(ei)}$. The electron mobility is
given then by
\begin{equation}
\widetilde{\mu }=\frac{e}{m\widetilde{\nu }}.  \label{meff}
\end{equation}

Substituting the interaction Hamiltonians, given by Eqs. (\ref{3}) and (\ref%
{4}), into Eq. (\ref{27}) one obtains

\begin{equation}
\widetilde{\nu }^{(eg)}=\frac{3\pi \hslash ^{4}\gamma a_{0}^{2}n_{g}}{%
8m^{3}T}\int_{0}^{\infty }q^{3}S(q,0)dq;  \label{29}
\end{equation}%
\begin{eqnarray}
\widetilde{\nu }^{(er)} &=&\frac{1}{4\pi mT}\int_{0}^{\infty }q^{3}[\frac{%
\hslash q\tanh (qd)}{2\rho \omega _{q}}]  \nonumber \\
&&\times |\langle \chi _{1}|V_{rq}(z)|\chi _{1}\rangle
|^{2}N_{q}S(q,\omega _{q})dq.  \label{30}
\end{eqnarray}%
The calculation of $\widetilde{\nu }^{(ei)}$ is done in a similar
way to
that of $\widetilde{\nu }^{(eg)}$ and the result is straightforward%
\begin{eqnarray}
\widetilde{\nu }^{(ei)} &=&\frac{1}{4\pi mT}\int_{0}^{\infty }q^{3}|\xi _{2%
\mathbf{q}}|^{2}|  \nonumber \\
&&\times \langle \chi _{1}|V_{iq}(z)|\chi _{1}\rangle
|^{2}S(q,0)dq. \label{31}
\end{eqnarray}

In the BEFM approach, the form factor $S(q,\omega )$\ is essential
to the evaluation of $\widetilde{\mu }$. Unfortunately $S(q,\omega
)$ can be calculated analytically only in the case of the
noninteracting electron system and there are no reliable
approximate expressions for $S(q,\omega )$ in the whole frequency
range in contrast to the static structure factor $S(k) $ $\neq
S(k,0)$ which can be evaluated by appropriate approximation
methods. In our calculations, we replace $S(q,\omega )$ in Eqs.
(\ref{29}-\ref{31})
by the noninteracting dynamical structure factor\cite{studart80,totsuji80}%
\begin{equation}
S(q,\omega )=\left(\frac{2\pi m}{Tq^{2}}\right)^{1/2}\exp \left(-\frac{\hslash ^{2}q^{2}}{8mT%
}-\frac{m\omega ^{2}}{2Tq^{2}}+\frac{\hslash \omega }{2T}\right)
\label{32}
\end{equation}%
and consider the limit $\hslash \omega _{q}/T\ll 1$.

\section{Results and Discussion}

Now we present the results of numerical calculations for SE
mobility over a helium film from Eqs. (\ref{9}) and (\ref{meff}).
We have used the parameters $\xi _{0}$ and $l$ of the
electron-interface scattering to have the best fit to the mobility
experimental data. First we must say that the mobility curves in
the BEA and FBEM exhibit the same overall behavior independently
of the adjustable parameters for the electron-interface
interaction. Results for the mobility dependence on the film
thickness in the BEA are presented in Fig. 1 for a glass substrate
$(\varepsilon _{s}=7.3) $ for $T=1.5$ K and the fitting parameters
$\xi _{0}=10$ \AA\ and $l=1420$ \AA . For a sake of comparison, we
plot the mobility data taken from Ref. \onlinecite{hucarmidahm92}.
One can see the decisive role of electron-interface scattering and
the best fit is attainable for $d\lesssim 300$ \AA , where, as
expected, the influence of roughness is more pronounced. We also
show separately the contributions to the mobility coming from
electron-ripplon and electron-atom scattering processes. We
observe that neither electron-gas nor electron-ripplon scattering
mechanisms can explain the experimental data for wide range of
$d$. By fitting the experimental data with mobility calculations
in the FBEM, we obtain the best results for $\xi _{0}=6.45$ \AA\
and $l=5000$ \AA .

\begin{figure}[tbp]
\includegraphics*[width=1.0\linewidth]{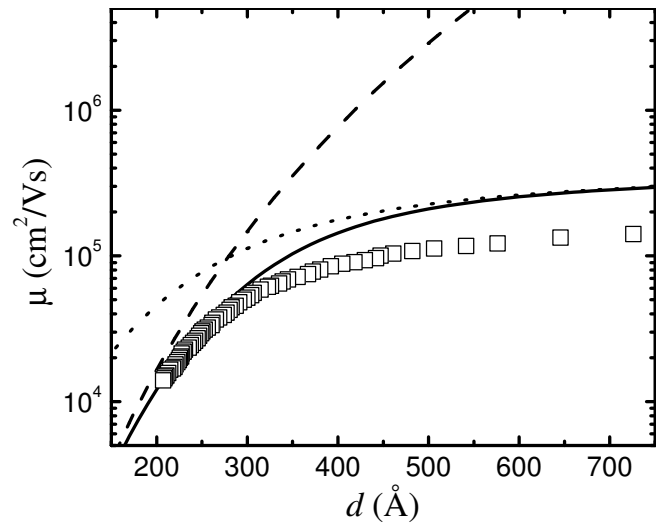}
\caption{Surface electron mobility as a function of the film
thickness for the helium film over a glass calculated within the
BEA. The solid line represents the general result when all
scattering processes are taken into account. The dashed line is
the contribution from interface roughness scattering and the
dotted line is the sum from contributions of the electron-ripplon
and electron-gas scattering. The squares are experimental points
of Ref. \onlinecite{hucarmidahm92}.}
\end{figure}

In Fig. 2 we depict, in a similar way to Fig. 1, mobility curves
as a
function of the film thickness for solid neon $(\varepsilon _{s}=1.19)$ and $%
T=1.2$ K. The best fit of experimental data \cite{platzman86} is
achieved
for $\xi _{0}=180$ \AA\ and $l=300$ \AA\ in BEA and $\xi _{0}=85$ \AA\ and $%
l=300$ \AA\ in FBEM. We see now that $\xi _{0}$ and $l$ are of the
same order of
magnitude of $d$ which makes unjustifiable the approximation $|\xi _{2}(%
\mathbf{r})|/d\ll 1$ which supports the perturbative approach to
obtain the potential $V_{iq}(z)$ of Eq. (\ref{4}), and, hence, the
applicability of Born approximation for the description of
electron-interface scattering. We would guess that the neon
surface, being much rougher than the glass one, it has more
irregularities leading to significant fluctuations of $d.$
Moreover, we observe a worse agreement between the theoretical
curve and experimental data at large $d$, where the
electron-ripplon and electron-gas mechanisms would dominated the
scattering.

\begin{figure}[tbp]
\includegraphics*[width=1.0\linewidth]{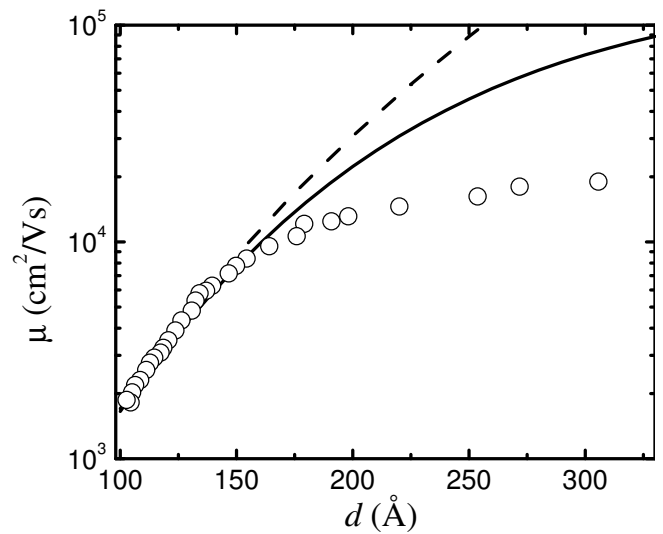}
\caption{Same as in Fig. 1 but for a solid neon substrate. The
solid line is the contribution coming from all scattering
processes and dashed line represents the mobility when only the
electron-interface roughness scattering is considered. The
experimental points are taken from Ref. \onlinecite {platzman86}.}
\end{figure}

The temperature dependence of SE mobility is depicted in Fig. 3
where the same values for $\xi _{0}$ and $l$ obtained by fitting
$\mu (d)$ are used. As it can be
seen we get rather good agreement with the experimental data of Ref. \onlinecite%
{jiangstandahm88} for $d=350$ \AA . However the agreement becomes
less satisfactory for smaller $d$. By considering only the
electron-ripplon and electron-atom scattering we found a
difference of about more than one order of magnitude between the
calculation results and the experimental ones. Unfortunately, to
our knowledge, there is no experimental data on the temperature
dependence of SE mobility in the case of a film over solid neon.

\begin{figure}[tbp]
\includegraphics*[width=1.0\linewidth]{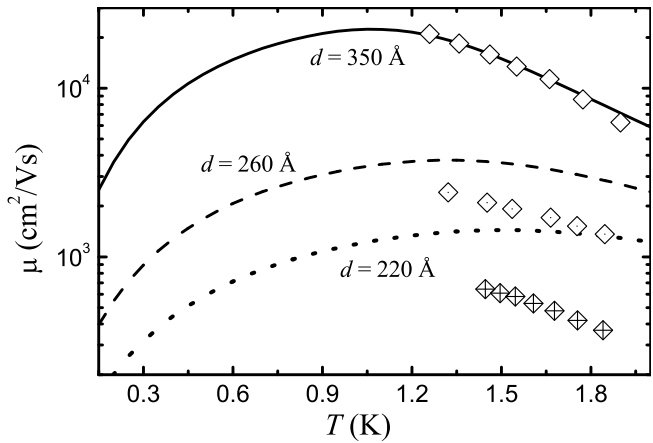}
\caption{SE mobility as a function of temperature including all
scattering processes over a glass substrate for some values of the
film thickness. The marks correspond to experimental data taken
from Ref. \onlinecite{jiangstandahm88}.}
\end{figure}

Figure 4 shows the influence of the dielectric constant of the
substrate on
the SE mobility calculated within the BEA. We used here the same values of $%
\xi _{0}$ and $l$ as in Fig. 1 for both glass and pmma
$(\varepsilon _{s}=2.2)$
substrates. We see that the mobility varies inversely with respect to $%
\varepsilon _{s}$ for whole of temperatures considered. This is a
direct consequence from $\Lambda _{1}\sim \varepsilon _{s}$
appearing in the collision frequencies which in turn are in the
denominator of the mobility formulas [Eqs. (\ref{9}) and
(\ref{meff})].

\begin{figure}[tbp]
\includegraphics*[width=1.0\linewidth]{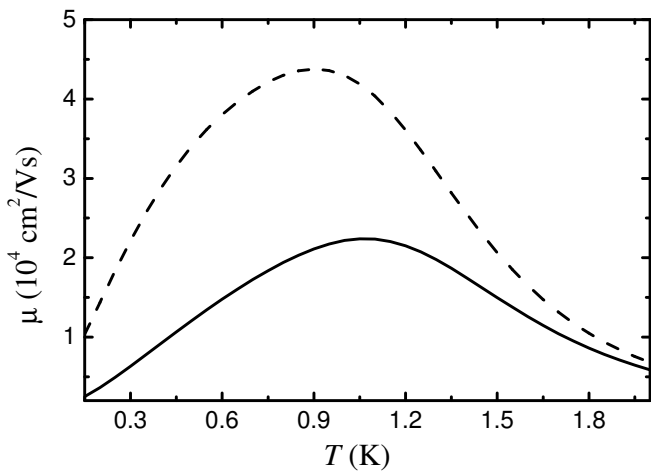}
\caption{Mobility versus temperature for a glass (solid line) and
pmma (dashed line) substrates.}
\end{figure}

In order to make clearer the role of different scattering
mechanisms we present in Fig. 5 the SE mobility, calculated within
the FBEM, as a function of temperature for a film over glass and
$d=350$ \AA . The calculated mobility curves are shown when the
distinct scattering processes are considered separately. As it can
be seen, the inclusion of roughness interface scattering is
essential to explain the experimental data. The nonmonotonic
dependence of SE mobility must be attributed to the crucial role
of electron-interface scattering since it is well known that the
mobility limited only by electron-ripplon and electron-gas
scattering is a monotonous decreasing function of temperature.

\begin{figure}[tbp]
\includegraphics*[width=1.0\linewidth]{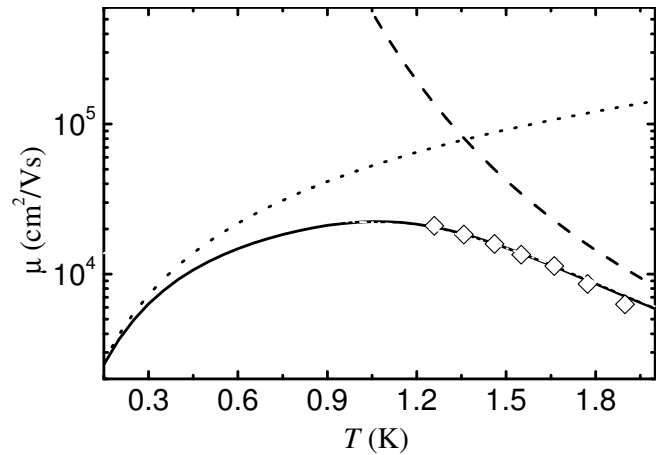}
\caption{Mobility as a function of the temperature calculated
within the FBEM for $d=350$ \AA\ and glass substrate. The solid
line is the contribution from all scattering mechanisms whereas
the dashed line is the contribution of gas scattering and the
dotted line is the result for only the interface roughness
scattering process.}
\end{figure}

\begin{figure}[tbp]
\includegraphics*[width=1.0\linewidth]{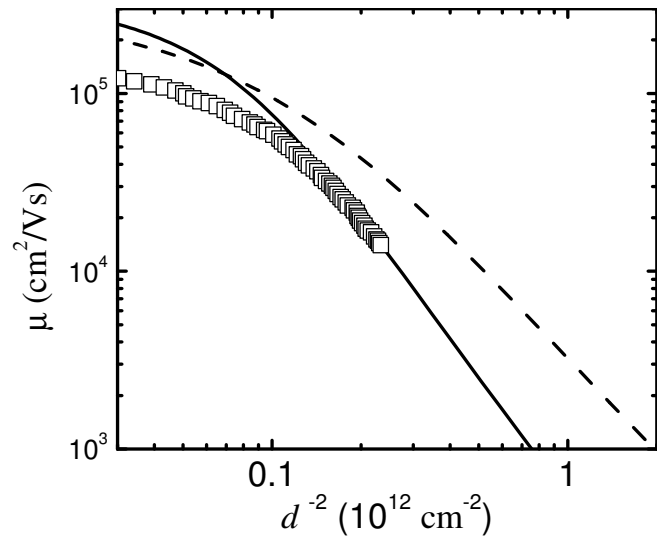}
\caption{Mobility as a function of the film thickness within the
FBEM (solid line) and BEA (dashed line) for a glass substrate.}
\end{figure}

\begin{figure}[tbp]
\includegraphics*[width=1.0\linewidth]{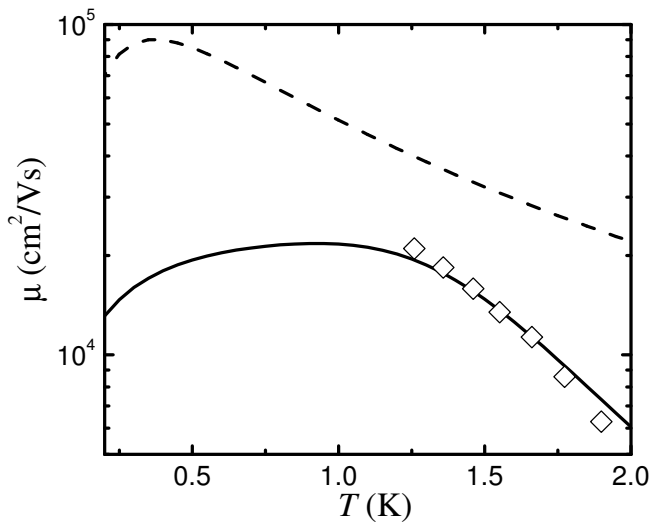}
\caption{Mobility as a function of temperature. The notations are
the same as in Fig. 6.}
\end{figure}

Finally, we compare the mobility results obtained in the two
different methods used here by depicting the curves of $\mu (d)$
and $\mu (T)$ in Figs. 6 and 7 respectively for the case of a
glass substrate. We plot also with the experimental data. \
Smaller values of $\xi _{0}$ obtained in FBEM
as compared with those in BEA fulfill more satisfactorily the condition $%
|\xi _{2}(\mathbf{r})|/d\ll 1$. We observe much better agreement
between experimental data and theoretical curves calculated in
FBEM for the same values of $\xi _{0}$ and $l$. We point out that
the numerical results within the FBEM are the same as those
obtained in the complete control approximation
(CCA).\cite{debora02,sokolov05} In the CCA, the mobility is
calculated with the BEA, but in the regime where the
electron-electron collision frequency is much higher than the
collision frequency due to other mechanisms. The electrons are
supposed to redistribute their momenta, due to collisions between
electrons, in such a way that the momentum of the total system
does not change but electrons acquire the same drift velocity. The
coincidence of results indicates that in order to establish the
specific role of electron-electron interactions we must go beyond
the noninteracting dynamical structure factor for calculating the
mobility in the FBEM as given by Eq. (\ref{32}). Our conclusions
then are valid only in the low or intermediate electron density
regime.

\section{Concluding Remarks}

In this work we have studied the dependence of the mobility of
surface electrons over a liquid helium film on the temperature and
on the film thickness. The transport properties are determined
from the solution of the Boltzmann equation in the relaxation time
approximation and in the force balance equation. We employed the
Gaussian correlated model to describe the interface roughness
between the film and the substrate. The parameters are adjustable
through the evaluation of both temperature and film thickness
dependencies of the mobility. The values obtained for the strength
and range of the rough interface are of the order of few atomic
layers and $100$ \AA\ respectively which are reliable and in the
mesoscopic regime. We showed that the SE mobility is limited by
the roughness scattering because electron-ripplon and electron-gas
atom scattering cannot explain the experimental data. However, we
cannot rule out the presence of other mechanisms as SE
localization by potential wells caused by the underlying solid
substrate.\cite{shikin01}

\section{Acknowledgment}

We are indebted to Professor A. J. Dahm, whom forthwith provides
us values of specific constants. This work was supported by CNPq
and FAPESP, Brazil.

\end{document}